\documentstyle[12pt]{article}
\begin{document}
\def\be{\begin{equation}}
\def\ee{\end{equation}}
\def\bea{\begin{eqnarray}}
\def\eea{\end{eqnarray}}
\def\lsim{\:\raisebox{-0.5ex}{$\stackrel{\textstyle<}{\sim}$}\:}
\def\gsim{\:\raisebox{-0.5ex}{$\stackrel{\textstyle>}{\sim}$}\:}

\begin{flushright}
TIFR/TH/93-59 \\
\end{flushright}
\bigskip
\bigskip
\begin{center}
\large {\bf Deterministic Quantum Mechanics} \\
\bigskip
\bigskip
\bigskip
S.M. Roy and Virendra Singh \\
\bigskip
Theoretical Physics Group \\
Tata Institute of Fundamental Research \\
Homi Bhabha Road, Bombay 400 005, India \\
\end{center}
\bigskip
\bigskip
\bigskip

\baselineskip=.85cm
\noindent \underbar{\bf Abstract}: Present quantum theory, which is
statistical in nature, does not predict joint probability
distribution of position and momentum because they are
noncommuting.  We propose a deterministic quantum theory
which predicts a joint probability distribution such that
the separate probability distributions for position and momentum agree
with usual quantum theory.  Unlike the Wigner distribution the suggested
distribution is positive definite.  The theory predicts a correlation
between position and momentum in individual events.

\bigskip
\bigskip
\bigskip
\bigskip

\hrule width 6cm
\bigskip

\noindent PACS : 03.65.Bz

\newpage

\noindent {\bf 1. \underbar{\bf Introduction}}.  Present quantum theory
does not
make definite prediction of the value of an observable in an individual
observation except in an eigenstate of the observable.  Application of
quantum rules to
two separated systems which interacted in the past together with a local
reality principle (Einstein locality) led Einstein, Podolsky and Rosen$^1$
to conclude that quantum theory is incomplete.  Bell$^2$ showed that
previous proofs of impossibility of a theory more complete than quantum
mechanics$^3$ (inappropriately called hidden variable theory) made
unreasonable assumptions; he went on however to prove$^4$ that a hidden
variable theory agreeing with the statistical predictions of quantum
theory cannot obey Einstein locality.

Bell's research was influenced by the construction by De Broglie and
Bohm$^5$ (dBB) of a hidden variable theory which reproduced the position
probability density of quantum mechanics but violated Einstein locality
for many particle systems.  For a single particle moving in one dimension
with Hamiltonian
$$
H = - \hbar^2/(2m) \partial^2/\partial x^2 + U(x),
\eqno (1)
$$
and wave function $\psi (x,t)$, de Broglie-Bohm proposed the complete
description of the state to be $\{\lambda (t),|\psi\rangle\}$, where
$\lambda (t)$ is the instantaneous position of the particle, and its
momentum is
$$
\hat p_{dBB} (\lambda,t) = md\lambda/dt = [{\rm Re}~\psi^\star (-i\hbar
{}~\partial \psi/\partial x)/(|\psi|^2)]_{x=\lambda}.
\eqno (2)
$$
In an ensemble the position density $\rho(\lambda,t)$ agrees with
$|\psi (\lambda,t)|^2$ for all time.  However,
Takabayasi$^6$ pointed out that the joint probability distribution
for position and momentum given by the theory
$$
\rho_{\rm dBB} (\lambda,p,t) = |\psi(\lambda,t)|^2 \delta\left(p - \hat
p_{dBB}(\lambda,t)\right)
\eqno (3)
$$
does not yield the correct quantum mechanical expectation value of $p^n$
for integral $n \neq 1$.  De Broglie$^5$ stated that these values in his
theory ``correspond to the
unobservable probability distribution existing prior to any
measurement'' and measurement will reveal different values distributed
according to standard statistical quantum mechanical formula.  Such a
central role for measurement is unsatisfactory if one wishes to apply the
dBB theory to closed quantum systems.

Without using hidden variables, Griffiths$^7$ and Gell-Mann and Hartle$^8$
introduced joint probability distributions for noncommuting observables at
different times in the consistent history approach to quantum theory of
closed systems.  Wigner$^9$ had earlier introduced a joint distribution
for $x$ and $p$ at the same time,
$$
\rho_W (x,p,t) = \int^\infty_{-\infty} {dy \over 2\pi\hbar} \psi^\star
\left(x + {y \over 2},t\right) \psi\left(x - {y \over 2},t\right)
\exp(ipy/\hbar)
\eqno (4)
$$
which yielded the correct quantum probability distributions separately for
$x$ and $p$ on integration over $p$ and $x$ respectively.  The Wigner
distribution cannot however be considered a probability distribution
because it is not positive definite, as seen from the fact that the
integral
$$
\int dx~dp~\rho_{W,\psi} (x,p) \rho_{W,\phi} (x,p) =
|(\psi,\phi)|^2/(2\pi\hbar)
$$
vanishes for two orthogonal states $\psi,\phi$.

We wish now to propose a deterministic quantum theory of a closed system
with the following properties.  (We consider in this paper only 1 particle
in 1 space dimension).

\noindent (i) At each time, the particle has a definite position
$x$ and a definite momentum $p$.

\noindent (ii) The system point in phase space has a Hamiltonian flow with
a $c$-number causal Hamiltonian $H_C (x,p,\psi(x,t),t)$ so that in an
ensemble of mental copies of the system the phase space density
$\rho(x,p,t)$ obeys Liouville's theorem
$$
d\rho (x,p,t)/dt = 0.
\eqno (5)
$$
Here $\psi(x,t)$ is the solution of the usual Schr\"odinger equation
$$
i\hbar~\partial \psi(x,t)/\partial t = H\psi(x,t)
\eqno (6)
$$
with $H$ being the standard quantum mechanical Hamiltonian for the system
and $H_C$ being determined from the following criteria.

\noindent (iii) Each pure ``causal state'', i.e., a set of phase space
points moving according to a single causal Hamiltonian $H_C$ has phase
space density of the deterministic form
$$
\rho(x,p,t) = |\psi(x,t)|^2 \delta\left(p - \hat p(x,t)\right),
\eqno (7)
$$
in which $p - \hat p(x,t) = 0$ not only determines $p$ as a function of
$x$, but also determines $x$ as a function of $p$ at each time (step
functions being allowed when necessary).  Eqn. (7)
guarantees on integration over $p$ the correct quantum probability
distribution in $x$ for any real function $\hat p(x,t)$.  The function is
determined from the requirement that on integration over $x$,
$\rho(x,p,t)$ should also yield the correct quantum probability distribution in
$p$.  That such a determination is possible and unique apart from a
discrete 2-fold ambiguity will be a crucial part of the present theory.
It is obvious that our $\hat p(x,t)$ will have to be different from the
$\hat p(x,t)$ of de Broglie-Bohm theory.

\noindent (iv) Since the quantum probability distributions for $x$ and $p$
in the statistics of many measurements are exactly reproduced, so are the
standard uncertainty relations.  However, the correlation between position
and momentum in individual events given by $\hat p(x,t)$ is an additional
testable prediction of the present theory.

In Secs. II, III we describe the construction of the momentum $\hat
p(x,t)$ and the causal Hamiltonian $H_C$, in Sec. IV applications to simple
quantum systems, and in Sec. V conceptual features of the new mechanics.
\bigskip

\noindent {\bf 2. \underbar{\bf Construction of Joint Probability
Distribution of position and}}

\underbar{\bf momentum.}  We seek a positive definite distribution of the
form (7) where
$\hat p$ is a monotonic function of $x$
$$
\epsilon~\partial\hat p (x,t)/\partial x \geq 0, ~~~~~ \epsilon = \pm 1
\eqno (8)
$$
The monotonicity
property ensures that for a given $t$, the $\delta$-function establishes a
one-to-one invertible correspondence between $x$ and $p$ whenever
$\partial \hat p/\partial x$ is finite and non-zero.
The requirement of reproducing the correct quantum probability
distribution of $p$ is that
$$
\int^\infty_{-\infty} \rho(x,p,t)dx = {1 \over \hbar} \big|\tilde
\psi \left({p \over \hbar},t\right)\big|^2,
\eqno (9)
$$
where $\tilde \psi (k,t)$ is the Fourier transform of $\psi(x,t)$.  We
substitute the ansatz (7) into (9) and integrate over momentum to obtain
$$
\int^p_{-\infty} dp' \int_{\hat p (x',t) \leq p} dx' |\psi(x',t)|^2
\delta\left(p' - \hat p(x',t)\right) = \int^p_{-\infty} {dp' \over \hbar}
\big|\tilde \psi\left({p' \over \hbar},t\right)\big|^2.
\eqno (10)
$$
The region $\hat p(x',t) \leq p$ becomes $x' \leq x$ if $\epsilon = 1$,
and $x' \geq x$ if $\epsilon = -1$, where $\hat p(x,t) = p$.  Thus, we
obtain, for $\epsilon = \pm 1$,
$$
\int^{\epsilon x}_{-\infty} dx' |\psi(\epsilon x',t)|^2 = \int^{\hat
p(x,t)/\hbar}_{-\infty} dk' |\tilde\psi (k',t)|^2.
\eqno (11)
$$
The left-hand side is a monotonic function of $x$ which tends to 1 for
$\epsilon x \rightarrow \infty$ for a normalized wave function; the
right-hand side
is a monotonic function of $\hat p$ tending to 1 for $\hat p \rightarrow
\infty$ (Parseval's theorem).  Hence, for each $t$, Eq. (11)
determines two monotonic functions $\hat p$ of $x$, one for each sign of
$\epsilon$.  (Note that the curve $\hat p (x,t)$ may have segments parallel
to $x$-axis or $p$-axis corresponding to $\psi(x,t)$ or
$\tilde\psi(p/\hbar,t)$ vanishing in some segment).  The two curves $p =
\hat p_\pm (x,t)$ so determined yield via Eq. (7) phase space densities
$\rho_\pm$, with different causal Hamiltonians $(H_C)_\pm$ determined below.

\bigskip

\noindent {\bf 3. \underbar{\bf Determination of the Causal Hamiltonian.}}
We view $\rho(x,p,t)$ as describing an ensemble of
system trajectories in the phase space.  We saw in the last section that
such a description is possible at each time.  We would now like to find a
causal Hamiltonian such that the time evolution in phase space implied
thereby is consistent with the time dependent Schr\"odinger equation.

In order that the total number of trajectories is conserved in time we
must have the continuity equation
$$
\partial \rho/\partial t + \partial (\rho
\dot x)/\partial x  + \partial (\rho \dot p)/\partial p  = 0
\eqno (12)
$$
If the dynamics of the trajectories is of Hamiltonian nature i.e.
$$
\dot x = \partial H_C/\partial p, ~~~
\dot p = -\partial H_C/\partial x
\eqno (13)
$$
then we have Liouville's theorem that the phase space density is conserved,
$$
\partial \rho/\partial t + \dot x \partial \rho/\partial
x + \dot p \partial \rho/\partial p = 0
\eqno (14)
$$
i.e.
$$
\partial \rho/\partial t + (\partial H_C/\partial p)
{}~\partial \rho/\partial x - (\partial H_C/\partial x)
{}~\partial \rho/\partial p = 0.
\eqno (15)
$$
The $c$-number Hamiltonian $H_C$ describing the causal time evolution of
the trajectories in the phase space will be allowed to be different
from the usual $q$-number Hamiltonian $H$ describing the time evolution of
the Schr\"odinger wave function $\psi$ according to Eq.
(6).

On substituting into Eq. (15) the ansatz (7)
discussed in the last section, we obtain
$$
\xi \delta(p - \hat p) + {\partial \over \partial
p} \left(\eta \delta(p - \hat p)\right) = 0
\eqno (16)
$$
where
$$
\begin{array}{l}
\xi = \partial |\psi|^2/\partial t +
(\partial H_C/\partial p) ~\partial |\psi|^2/\partial x -
\partial \eta/\partial p \\[2mm]
\eta = -|\psi|^2 \left\{\partial \hat p/\partial t +
(\partial \hat p/\partial x) ~\partial
H_C/\partial p + \partial H_C/\partial x\right\}.
\end{array}
$$

We thus need for consistency
$$
\xi = 0 ~~~{\rm and}~~~ \eta = 0 ~{\rm if}~ p = \hat p.
\eqno (17)
$$
We now specialise to the usual case when H is given by Eq. (1).
We find that this situation is taken care of with the choice of $H_C (x,
p, t)$
$$
H_C = {1 \over 2m} (p - A(x,t))^2 + V(x,t).
\eqno (18)
$$
The causal Hamiltonian is of the Newtonian form apart from the
introduction of a vector potential $A(x,t)$ and allowing the potential
$V(x,t)$ to differ from $U(x)$.  Eqs. (17) lead to the following equations
to determine $V$ and $A$ (after using Schr\"odinger eqn. to substitute for
$\partial |\psi|^2/\partial t$),
$$
- \partial V(x,t)/\partial x = \partial \hat p(x,t)/
\partial t + (2m)^{-1} \partial\left(\hat p(x,t) -
A(x,t)\right)^2/\partial x,
\eqno (19)
$$
$$
\partial \left[|\psi|^2 (\hat p - A - mv)\right]/\partial x = 0,
\eqno (20)
$$
where $v$ is given by
$$
v(x,t) = \hbar/(2im) ~\partial\ell n (\psi/\psi^\star)/\partial x
\eqno (21)
$$
which is just the de Broglie-Bohm velocity.  Eq. (20) implies that the
quantity in square brackets must be a function of $t$ alone.  We choose
this function of $t$ to be zero in order to avoid a singularity of the
vector potential at the nodes of the wave function.  We thus obtain
$$
A(x,t) = \hat p(x,t) - mv(x,t)
\eqno (22)
$$
With the calculation of the causal Hamiltonian thus completed via Eqs.
(18), (19) and (22) a consistent Liouville description emerges.

\bigskip

\noindent {\bf 4. \underbar{\bf Illustrative Examples.}} (i) Quantum Free
Particle.
Let the quantum free particle be described by the Gaussian momentum space
wave function
$$
\tilde \psi(p/\hbar,t) = (2\pi)^{-1/4}
\exp\left[-(p-\beta)^2/(2\alpha\hbar^2) - ip^2t/(2m\hbar)\right]
\eqno (23)
$$
so that the coordinate space wave function is
$$
\psi (x,t) =  (\pi \alpha)^{-1/4}
\left(m \alpha/(m + i \alpha \hbar t)\right)^{1/2} \exp
f,
\eqno (24)
$$
$$
f = - (\alpha/2)
\left[(x - \beta t/m)^2 - i\left(\displaystyle{\alpha \hbar t \over m} x^2
+ \displaystyle{2\beta x \over \alpha \hbar} - \displaystyle{\beta^2 t
\over m \alpha \hbar}\right)\right]\Big/\left(1 + \displaystyle{\alpha^2
\hbar^2 t^2 \over m^2}\right).
$$
Our procedure yields
$$
A = \hat p - \beta = \pm \hbar \sqrt{{m^2 \alpha^2 \over m^2 + \alpha^2
\hbar^2 t^2}} \left(x - {\beta t \over m}\right),
\eqno (25)
$$
and
$$
\partial V/\partial x = \pm  (m^2 + \alpha^2\hbar^2
t^2)^{-3/2} [xt(\alpha\hbar)^2 + \beta m] (\hbar\alpha m)
\eqno (26)
$$
The determination of the causal Hamiltonian is now complete apart from an
irrelevant additive function of $t$.  Eq. (25) apart from predicting the
momentum $\hat p = \beta$ at the centre of the wave packet $x = \beta
t/m$, also predicts values of $\hat p$ at other values of $x$ which the
particle may have in individual events.  The quantum potentials $A$ and
$V$ are seen to be proportional to $\hbar$ in this example.

\noindent (ii) Quantum Oscillator.  For the minimum uncertainty coherent
state of the harmonic oscillator of mss $m$, angular frequency $\omega$ and
amplitude of oscillation $a$ we find
$$
\rho(x,p,t) = \sqrt{{m\omega \over \pi\hbar}} \exp\left[-{1\over2}
{m\omega \over \hbar} \left(x - a \cos (\omega t)\right)^2\right]
\delta\left(p -\hat p(x,t)\right),
\eqno (27)
$$
where
$$
\hat p(x,t) = -m\omega a \sin(\omega t) \pm m\omega(x-a\cos(\omega t)),
\eqno (28)
$$
and
$$
A(x,t) = \pm m\omega (x - a \cos(\omega t)),
\eqno (29)
$$
$$
- \partial V(x,t)/\partial x = -m\omega^2 a \cos(\omega t) \pm
m\omega^2 a \sin (\omega t)
\eqno (30)
$$
The causal Hamiltonian yields the equation of motion
$$
md^2 x/dt^2 = -m\omega^2 a \cos \omega t
\eqno (31)
$$
which results in exact harmonic motion even for $x$ away from the centre
of the packet.  We do not of course expect this for solutions of the
Schr\"odinger eqn. different from the coherent state here considered.

\bigskip

\noindent {\bf 5. \underbar{\bf New conceptual features.}} (a) We have
derived corresponding to every quantum wave function
$\psi$, two joint probability distributions for position and momentum of
the form (7) which are (i) positive definite, (ii) have Hamiltonian
evolution with causal Hamiltonians $(H_C)_\pm$ and obey (iii)
$$
\int (f(x) + g(p)) \rho_\pm(x,p,t) dx~dp = \left(\psi,\left(f(x) +
g\left(-i\hbar {\partial \over \partial x}\right)\right)\psi\right),
\eqno (32)
$$
for arbitrary functions $f(x)$ and $g(p)$.  Eq. (32) is the major
advantage of the present theory over the $dBB$ theory.  It raises the
exciting possibility that the momentum values $\hat p_\pm (x,t)$ for
individual events here derived could agree with experimental values, and
the single particle theory described could be the seed of a general
quantum theory of closed systems.  We postpone the discussion of
measurements until we present a generalization of the theory to many
particles.

\noindent (b) Since both $\rho_+$ and $\rho_-$ obey Eq. (32) so will $\rho
= C\rho_+ + (1-C)\rho_-$ with $0 \leq C \leq 1$.  But since $\rho_+$ and
$\rho_-$ correspond to different causal Hamiltonians $(H_C)_\pm$, $\rho$
will not correspond to a `pure causal state'.  We are led to the concept
of a pure causal state as being more fine grained than a pure wave
function $\psi$.  All $\rho = C\rho_+ + (1-C)\rho_-$ correspond to $\psi$
$(\rho \leftrightarrow \psi)$ for a continuum of values of $C$, but only
$C = 0,1$ correspond to pure causal states.  To quantum density matrix
states $\Sigma C_\alpha |\psi_\alpha\rangle \langle \psi_\alpha|$
correspond phase space densities $\Sigma C_\alpha \rho_\alpha$ if
$\rho_\alpha \leftrightarrow \psi_\alpha$.

\noindent (c) It is clear that the causal Hamiltonian evolution of phase
space densities could be described purely in the phase space language
without using the intermediate step of the wave function.  We find it
convenient to use $\psi(x,t)$ at the present stage.

\noindent (d) One can ask if Eq. (32) can be generalized to more general
quantum observables.  Here we face the old problem that there exist
nonclassical observables e.g. $x\left(-i\hbar {\partial \over \partial
x}\right)x$, $\left(\left(-i\hbar {\partial \over \partial x}\right) xx +
h.c.\right)/2$ which have different expectation values but the same
`naive' classical analogue $x^2p$.  A trivial
way followed already for the $dBB$ distribution is: given a nonclassical
observable $A$ the phase space analogue can be $f(x,p,\psi)$ such that
$f(x,\hat p,\psi) = \psi^\star A\psi/|\psi|^2$.  Perhaps only those
operators $A$, which like $(f(x) + q(p))$ have a phase space
representation $f(x,p)$ independent of $\psi$ should be considered as
`beables'.$^{10}$

\noindent (e) The predictions of the momentum values $\hat
p(x,t)$ are independent of any special ansatz for $H_C$.

\noindent (f) Due to the existence of trajectories, the problem of
inconsistent histories will not arise in this theory.

John Bell's writings, especially his last article `against
measurement'$^{10}$ has profoundly influenced this work.  We are grateful
to colleagues at the Tata Institute especially
to Deepak Dhar for many stimulating questions.

\newpage

\noindent {\bf References}

\begin{enumerate}
\item A. Einstein, B. Podolsky and N. Rosen, Phys. Rev. \underbar{47}, 777
(1935).

\item J.S. Bell, Rev. Mod. Phys. \underbar{38}, 447 (1966).

\item E.g. J. Von Neumann, ``Mathematische Grundlagen der Quanten \break 
Mechanik'',
Julius Springer Verlag, Berlin (1932) (English Transl.: \break Princeton Univ.
Press, Princeton, N.J. 1955).

\item J.S. Bell, Physics (Long Island City, N.Y.) \underbar{1}, 195
(1964).

\item L. de Broglie, ``Nonlinear Wave Mechanics, a causal
interpretation'', (Elsevier,
Amsterdam 1960); D. Bohm, Phys. Rev. \underbar{85}, 166, 180 (1952); D.
Bohm, B.J. Hiley and P.N. Kaloyerou, Phys. Rep. \underbar{144}, 349
(1987);  P.R. Holland,  ``The Quantum Theory of Motion''
(Cambridge Univ. Press, Cambridge 1993).

\item T. Takabayasi, Progr. Theoret. Phys. \underbar{8}, 143 (1952).

\item R.B. Griffiths, J. Stat. Phys. \underbar{36}, 219 (1984); Phys. Rev.
Letts. \underbar{70}, 2201 (1993).

\item M. Gell-Mann and J.B. Hartle, in Proc. 25th Int. Conf. on High
Energy Physics, Singapore, 1990, Eds. K.K. Phua and Y. Yamaguchi (World
Scientific, Singapore 1991).  R. Omn\`es, Rev. Mod. Phys. \underbar{64},
339 (1992).

\item E. Wigner, Phys. Rev. \underbar{40}, 749 (1932).

\item J.S. Bell, `against measurement', in ``Sixty Two Years of
Uncertainty'', ed. A.I. Miller (Plenum, New York 1990).

\end{enumerate}

\end{document}